# Critical frequency in nuclear chiral rotation

P. Olbratowski,[1,2] J. Dobaczewski,[1,2] and J. Dudek[2]

[1]*Institute of Theoretical Physics, Warsaw University, Hoża 69, PL-00681, Warsaw, Poland*

[2]*Institut de Recherches Subatomiques, UMR7500, CNRS-IN2P3 and Université Louis Pasteur, F-67037 Strasbourg Cedex 2, France*

Within the cranked Skyrme-Hartree-Fock approach the self-consistent solutions have been obtained for planar and chiral rotational bands in $^{132}$La. It turns out that the chiral band cannot exist below some critical rotational frequency which in the present case equals $\hbar\omega$=0.6 MeV. The appearance of the critical frequency is explained in terms of a simple classical model of two gyroscopes coupled to a triaxial rigid body.

Spontaneous chiral symmetry breaking constitutes an interesting type of collective effect in fermion systems [1]. In the present context, a system is called chiral if it is not related to its time-reversed partner by a spatial rotation. A rotating triaxial body with its angular momentum pointing towards one of the octants of the intrinsic coordinate system, where the short ($s$), medium ($m$) and long ($l$) axes form either a left- or right-handed set, is an example. In the laboratory frame, the chiral rotation leads to pairs of rotational bands. Rigid bodies with no internal structure cannot rotate freely in such a way, and therefore, investigation of the chiral rotation may give us information on the underlying internal properties.

Triaxiality is common in molecules where it arises from a spatial arrangement of the constituent atoms. In nuclei it can be induced by shell effects. It has been suggested [1] that appearance of chiral doublets may provide a strong evidence for nuclear triaxiality. Candidates for such doublets were recently observed in several $N$=75 isotones [2], $^{132}$La among them. Calculations within the particle-rotor model [3] confirm their possible chiral character.

As proposed in Ref. [4], the chiral rotation may appear if a few proton particles occupy the lowest substates of a high-$j$ orbital, and simultaneously a few neutron holes are left in the highest substates. The former drive the nucleus towards elongated shapes, while the latter towards disc-like shapes. An interplay of these opposite tendencies may yield triaxiality. It also turns out that the considered particles and holes align their angular momenta along the short and long axes, respectively. As follows from hydrodynamical considerations [5], the moment of inertia along the medium axis is the largest, thus favoring the collective core rotation around this axis. Therefore, the total spin has non-zero components on all axes, creating the possibility of left- or right-handed rotation.

In the present study we consider the $\pi h_{11/2}^{1} \nu h_{11/2}^{-1}$ configuration in $^{132}$La, which fulfills the above prerequisite conditions for the chiral rotation. Our analysis is based on the Hartree-Fock (HF) cranking calculations [6] with the angular frequency vector that can arbitrarily orient itself with respect to the mass distribution (Tilted-Axis Cranking (TAC) model [7]). In all HF solutions found in $^{132}$La a stable triaxial deformation of about $\beta = 0.25$ and $\gamma = 45°$ was obtained. In the framework of the TAC model, with phenomenological mean field, chiral solutions were already obtained in $^{134}$Pr and $^{188}$Ir [8].

In order to study the rotational properties of the single-particle (s.p.) orbitals in question, we first applied the angular-momentum cranking in the $s$-$l$ plane for $0<\hbar\omega_s,\hbar\omega_l<0.5$ MeV. Calculations were performed non-selfconsistently, starting from the converged HF solution obtained at $\boldsymbol{\omega}$=0.

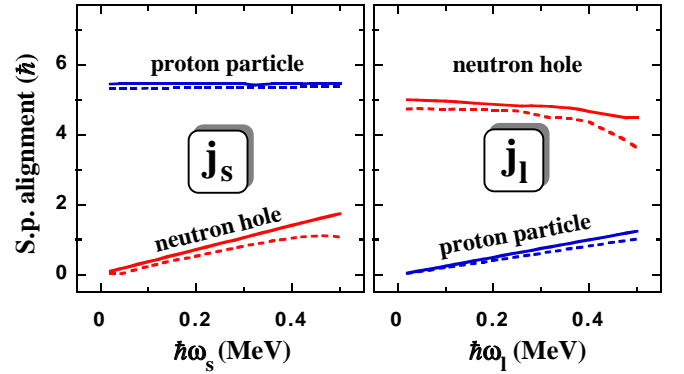

FIG. 1. Single-particle alignments on the short ($\boldsymbol{j}_s$) and long ($\boldsymbol{j}_l$) axes for the $h_{11/2}$ odd proton particle and neutron hole in $^{132}$La. Results of the non-selfconsistent cranking calculations are shown in function of $\omega_s$ (left) and $\omega_l$ (right). Solid lines correspond to $\hbar\omega_l$=0 (left) and $\hbar\omega_s$=0 (right), and dashed lines to $\hbar\omega_l$=0.5 (left) and $\hbar\omega_s$=0.5 MeV (right), while results for all intermediate values lie between the solid and dashed curves.

For $\boldsymbol{\omega}$=0 the proton particle does indeed align its spin along the short, and the neutron hole along the long axis, see Fig. 1. The response of these alignments to rotation is rather weak, meaning that the s.p. wave functions are strongly confined by deformation. In the proton case, only the alignment on the long axis changes in response to rotation and it depends only on $\omega_l$. The neutron align-ment shows similar properties with the short and long axes interchanged. These results can be summarized as follows

$$\boldsymbol{j}^p \simeq s_s\hat{\mathbf{i}}_s + \delta\mathcal{J}_l\omega_l\hat{\mathbf{i}}_l, \quad \boldsymbol{j}^n \simeq s_l\hat{\mathbf{i}}_l + \delta\mathcal{J}_s\omega_s\hat{\mathbf{i}}_s, \qquad (1)$$

where $\hat{\mathbf{i}}_s$ and $\hat{\mathbf{i}}_l$ are the unit vectors along the short and long axes, respectively. Therefore, the odd particle and hole behave as gyroscopes of spins $s_s$ and $s_l$, rigidly fixed along the long and short axes, respectively. In addition, the s.p. alignments also have small perpendicular components that amount to small contributions, $\delta\mathcal{J}_s$ and $\delta\mathcal{J}_l$, to the collective total moments of inertia.

Note, that the above analysis could not be done self-consistently, because we oriented vector $\boldsymbol{\omega}$ along all directions in the intrinsic frame. In self-consistent solutions its orientation is constrained by the Kerman-Onishi theorem [9] requiring that $\boldsymbol{\omega}$ and $\boldsymbol{I}$ be parallel.

To examine the collective rotational behavior we now turn to self-consistent calculations with the standard axial cranking, that is when both $\boldsymbol{\omega}$ and $\boldsymbol{I}$ point along either of the three principal axes of the mass distribution. Figure 2 shows the properties of the three resulting bands at frequencies of $\hbar\omega\leq 0.5$ MeV. One can see that at zero frequency cranking around the medium ($m$) axis gives a vanishing angular momentum, while those around the other two ($s$ and $l$) give non-zero values equal to the s.p. alignments of the odd proton particle and neutron hole, $s_s$=5.43 and $s_l$=5.00 $\hbar$, respectively. For each of the three bands the dependence of $I$ on $\omega$ is linear, like for a rigid rotation, and the corresponding slopes give the microscopic collective moments of inertia along the principal axes, $\mathcal{J}_s = 8.5$, $\mathcal{J}_m = 31$, and $\mathcal{J}_l = 20\,\hbar^2$/MeV. These values already contain the contributions, $\delta\mathcal{J}_s$ and $\delta\mathcal{J}_l$, from the odd particle and hole.

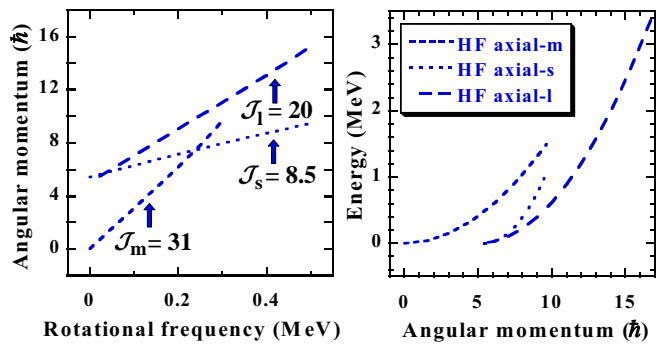

FIG. 2. Angular momenta (left) and excitation energies (right) in $^{132}$La are shown for the HF bands corresponding to the axial cranking along the short ($s$), medium ($m$), and long ($l$) principal axes.

The microscopic results presented up to now do show that the system can indeed be described in terms of two gyroscopes plus a collective core. It is therefore fully justified to use a particle-core coupling model to describe the resulting quantum states [3,4]. However, in the present study we also argue that basic features of the system are purely classical. To show that, we now discuss a very simple classical model that faithfully represents salient aspects of the underlying physical situation.

The model consists of a triaxial rigid body [10] with two gyroscopes rigidly fixed along its short and long axes, modeling the odd nucleons. Moments of inertia $\mathcal{J}_{s,m,l}$ obtained above are the diagonal components of the body inertia tensor $\hat{\mathcal{J}}$, which we otherwise assume diagonal, while $\boldsymbol{s} = s_s\hat{\mathbf{i}}_s + s_l\hat{\mathbf{i}}_l$ denotes the vector sum of the gyroscopes' spins.

The angular momentum of this system reads $\boldsymbol{I} = \hat{\mathcal{J}}\boldsymbol{\omega} + \boldsymbol{s}$. Here we look only for uniform rotations, by which we understand that $\boldsymbol{\omega}$ is constant in the body-fixed frame. In such a case, the Euler equations [11] take the form $\boldsymbol{\omega}\times\boldsymbol{I} = 0$, and require that $\boldsymbol{\omega}$ and $\boldsymbol{I}$ be parallel, like for the self-consistent cranking solutions [9]. These equations can easily be solved. However, to keep contact with the Hartree-Fock method, here we employ the variational principle to find the solutions.

The Lagrangian of the system is equal to the sum of kinetic energies of the body and gyroscopes [12],

$$L = E_{\text{kin}} = \tfrac{1}{2}\boldsymbol{\omega}\hat{\mathcal{J}}\boldsymbol{\omega} + \boldsymbol{\omega}\cdot\boldsymbol{s}, \qquad (2)$$

and for uniform rotations it is independent of time. Hence, minimization of the action integral $\int L dt$ is equivalent to minimizing the Lagrangian as function of $\boldsymbol{\omega}$.

Since the Hamiltonian of the system is $H = \boldsymbol{\omega}\cdot\boldsymbol{I} - L = \tfrac{1}{2}\boldsymbol{\omega}\hat{\mathcal{J}}\boldsymbol{\omega}$, the uniform solutions can be obtained by finding extrema of function $R = -L = H - \boldsymbol{\omega}\cdot\boldsymbol{I}$ which is called Routhian. This serves as a bridge between our classical model and the quantum cranking theory where an analogous Routhian is minimized in the space of trial Slater determinants.

Extrema of $R$ with respect to the components of $\boldsymbol{\omega}$ in the intrinsic frame at a given $|\boldsymbol{\omega}|$ can be found by using a Lagrange multiplier $\mu$ for $\omega^2$. Setting to zero the derivatives of the quantity:

$$R + \tfrac{1}{2}\mu\omega^2 = \tfrac{1}{2}[(\mu-\mathcal{J}_s)\omega_s^2 + (\mu-\mathcal{J}_m)\omega_m^2 \\ + (\mu-\mathcal{J}_l)\omega_l^2] - (\omega_s s_s + \omega_l s_l) \qquad (3)$$

with respect to $\omega_m$, $\omega_s$, and $\omega_l$, one gets

$$\omega_m(\mu - \mathcal{J}_m) = 0, \qquad (4a)$$
$$\omega_s = s_s/(\mu - \mathcal{J}_s), \qquad (4b)$$
$$\omega_l = s_l/(\mu - \mathcal{J}_l). \qquad (4c)$$

Equation (4a) gives either $\omega_m = 0$ or $\mu = \mathcal{J}_m$ leading to two distinct classes of solutions.

*Planar solutions:* If $\omega_m = 0$ then both $\boldsymbol{\omega}$ and $\boldsymbol{I}$ lie in the $s$-$l$ plane where also the gyroscopes' spins are located. This gives planar solutions for which the chiral symmetry is conserved. All values of $\mu$ are allowed, and the Lagrange multiplier must be determined from the magnitude of $\boldsymbol{\omega}$. Figure 3 shows $\omega$ versus $\mu$ for the parameters extracted from the HF solutions. As can be seen, there always exist two solutions and above some rotational frequency two more appear. This threshold frequency reads

$$\omega_{\text{thr}} = \frac{\left(s_s^{2/3} + s_l^{2/3}\right)^{3/2}}{|\mathcal{J}_l - \mathcal{J}_s|} \quad . \tag{5}$$

For the present model parameters, the threshold frequency is rather high, $\hbar\omega_{\text{thr}}=1.28\,\text{MeV}$.

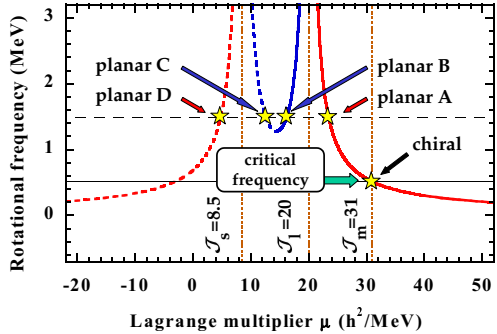

FIG. 3. Classical planar solutions for the uniform rotation are visualized as intersections of the horizontal line with the function $\omega(\mu)$ obtained from Eqs. (4bc). The classical chiral solution is given by $\mu=\mathcal{J}_m$.

*Chiral solutions:* For $\mu=\mathcal{J}_m$ all values of $\omega_m$ are allowed, while the components in the $s$-$l$ plane are fixed at

$$\omega_s = \frac{s_s}{\mathcal{J}_m - \mathcal{J}_s}, \quad \omega_l = \frac{s_l}{\mathcal{J}_m - \mathcal{J}_l}. \tag{6}$$

Consequently, the angular momentum has non-zero components along all three axes and the chiral symmetry is broken. For each value of $\omega$ there are two solutions differing by the sign of $\omega_m$, and thus giving a chiral doublet. The fixed non-zero values of $\omega_s$ and $\omega_l$ (6) lead to the principal conclusion that chiral solutions cannot exist for $\omega$ smaller than the critical frequency $\omega_{\text{crit}}$,

$$\omega_{\text{crit}} = \left[\left(\frac{s_s}{\mathcal{J}_m - \mathcal{J}_s}\right)^2 + \left(\frac{s_l}{\mathcal{J}_m - \mathcal{J}_l}\right)^2\right]^{1/2}. \tag{7}$$

In the present case we have $\hbar\omega_{\text{crit}}=0.51\,\text{MeV}$. Note, that for $\omega_m=0$ the chiral solution coincides with the planar A solution.

Figure 4 summarizes the angular momenta and energies for all the bands found in the classical rotor coupled to two gyroscopes. At low angular momenta, the yrast line coincides with the planar D band, where the core rotates in the opposite direction as compared to the gyroscopes. Then it continues along the planar A solution. Since the moment of inertia $\mathcal{J}_m$ is the largest, beyond the critical frequency the yrast line coincides with the chiral solution. This may give us hope of finding chiral bands in experiment.

We now come back to the microscopic self-consistent HF solutions. By applying rotational frequency with non-zero $s$ and $l$ components, we obtained a planar band corresponding to the classical A solution. At the limit of the zero frequency, the angular momentum is simply a sum of the contributions from the proton particle and neutron hole, that are almost equal. Therefore, the initial tilt angle is approximately equal to $45°$. With increasing $\omega$, the angular frequency vector $\boldsymbol{\omega}$ tilts more and more towards the long axis. At $\hbar\omega=0.75\,\text{MeV}$, $\boldsymbol{\omega}$ is tilted by $24°$. This HF evolution of the tilt angle is to a very high precision reproduced by our classical model (solid line in Fig. 5).

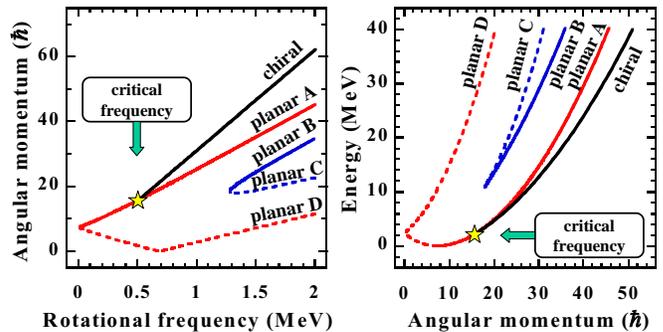

FIG. 4. Angular momenta $I(\omega)$ (left) and energies $E(I)$ (right) for the four planar and the chiral solutions obtained in the classical model.

To facilitate obtaining the chiral solution we made use of the classical prediction that it has a common point with the planar A solution. To each converged point of the planar band we applied a cranking frequency with a non-zero component along the medium axis. Chiral solution appeared from $\hbar\omega=0.6\,\text{MeV}$ on, which is in a qualitative agreement with the classical estimate of $\hbar\omega_{\text{crit}}=0.51\,\text{MeV}$. In spite of several attempts no chiral rotation was found below this value. This confirms the existence of a critical frequency within the Skyrme-Hartree-Fock method. The evolution of $\boldsymbol{\omega}$ along the HF chiral band is shown in Fig. 5 (triangles). There is again a qualitative agreement with the classical result.

We conclude the presentation of our results by comparing the HF bands with experimental data [2], Fig. 6. We first note that the HF critical frequency of $\hbar\omega_{\text{crit}}=0.6\,\text{MeV}$ is significantly higher than the frequencies at which appear the two experimental bands that are candidates for the chiral pair [2]. The classical model gives a clear dependence of $\omega_{\text{crit}}$ (7) on the parameters characterizing the system. Therefore, we may speculate that a smaller critical value can only be obtained by either making the s.p. alignments (spins of the gyroscopes) smaller, or the differences of the principal moments of inertia larger. Inclusion of pairing will probably leave $\boldsymbol{s}$ unchanged (blocked states), and at the same time decrease the moments of inertia, thus rendering $\omega_{\text{crit}}$ possibly even higher. Only making the deformation parameter $\gamma$ closer to 30°, could, according to the hydrodynamical model [5], decrease the differences of the diagonal moments of inertia and make $\omega_{\rm crit}$ lower. A case with $\gamma = 30°$ was considered in Ref. [4].

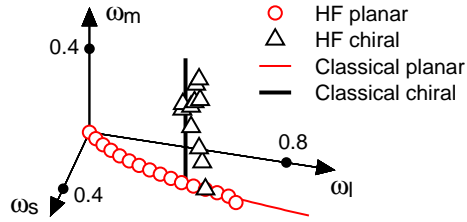

FIG. 5. Evolution of the angular frequency vector in the intrinsic frame along the HF planar (circles) and chiral (triangles) bands. Solid lines show the analogous bands obtained in the classical model. Scales are given in units of MeV/$\hbar$.

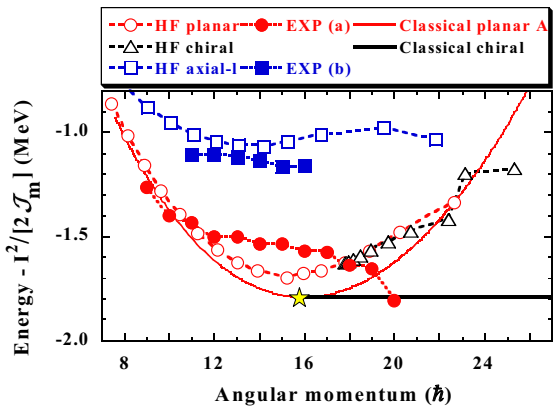

FIG. 6. Excitation energies of the calculated and experimental (yrast (a), side (b) [3]) bands in $^{132}$La. The classical rigid-rotor reference for $\mathcal{J}_m = 31\,\hbar^2/\text{MeV}$ is subtracted. Solid lines represent the results of the classical model for the chiral (thick line) and planar A solutions (thin line).

Chiral interpretation of the two bands in $^{132}$La is supported by the fact that standard axial cranking calculations systematically overestimate the ~350 keV splitting between them [2]. On the other hand, a possible alternative explanation might consist in associating the lower and higher band with the planar and axial-$l$ HF solutions, respectively. Both these bands correspond to $\Delta I=1$ sequences (two different proton signatures are possible for the axial-$l$ band), and the energy splitting of the two bands is well reproduced, see Fig. 6.

In summary, solutions corresponding to the planar and chiral nuclear bands were found for the first time within a fully self-consistent microscopic approach. We have shown that most of the characteristics of the HF solutions can be analyzed in terms of a simple model describing rotation of a triaxial body coupled to two gyroscopes. In particular, the model gives an analytic expression for the value of the critical angular frequency at which the chiral band begins. This value seems to be rather high as compared to the angular frequencies corresponding to the pair of experimental bands in $^{132}$La, where the existence of a chiral pair was suggested. A tentative interpretation of the experimental bands as those exhibiting a planar and axial-$l$ rotations was possible.

Interesting comments by W. Nazarewicz and K. Starosta are gratefully acknowledged. This research was supported in part by the Polish Committee for Scientific Research (KBN) under Contract No. 5 P03B 014 21, by the French-Polish integrated actions program POLONIUM, and by the computational grant from the Interdisciplinary Centre for Mathematical and Computational Modeling (ICM) of the Warsaw University.